\documentclass[preprint,showpacs,preprintnumbers,amsmath,amssymb]{revtex4}

\usepackage{graphicx}% Include figure files
\usepackage{dcolumn}% Align table columns on decimal point
\usepackage{bm}% bold math
\usepackage{color}

\begin{document}

\preprint{APS/123-QED}

\title{Phase transitions in nanoconfined binary mixtures of highly oriented colloidal rods}
\author{Daniel de las Heras} \email{daniel.delasheras@uam.es}
\affiliation{Departamento de F\'{\i}sica Te\'orica de la Materia Condensada,
Universidad Aut\'onoma de Madrid, E-28049 Madrid, Spain}
\author{Yuri Mart\'{\i}nez-Rat\'on}\email{yuri@math.uc3m.es}
\affiliation{Grupo Interdisciplinar de Sistemas Complejos (GISC),
Departamento de Matem\'{a}ticas,Escuela Polit\'{e}cnica Superior,
Universidad Carlos III de Madrid, Avenida de la Universidad 30, E--28911, Legan\'{e}s, Madrid, Spain}
\author{Enrique Velasco}\email{enrique.velasco@uam.es}
\affiliation{Departamento de F\'{\i}sica Te\'orica de la Materia Condensada
and Instituto de Ciencia de Materiales Nicol\'as Cabrera,
Universidad Aut\'onoma de Madrid, E-28049 Madrid, Spain}
\date{\today}

\begin{abstract}
We analyse a binary mixture of colloidal parallel hard cylindrical
particles with identical diameters but dissimilar lengths $L_1$ and $L_2$, with $s=L_2/L_1=3$,
confined by two parallel hard walls in a planar slit-pore geometry, using a fundamental--measure
density functional theory. This model presents (arXiv:1002.0612v, accepted in Phys. Rev. E) 
nematic (N) and two types of smectic (S) phases, with first- and second-order N-S bulk transitions and S-S demixing, 
and surface behaviour at a single hard wall which includes complete wetting by 
the S phase mediated (or not) by an infinite number of surface-induced layering (SIL) transitions. In the 
present paper the effects of confinement on this model colloidal fluid mixture are studied. 
Confinement brings about profound changes in the phase diagram, resulting from competition 
between the three relevant length scales: pore width $h$, smectic period $d$ and length ratio 
$s$. Four main effects are identified: (i) Second-order bulk N-S transitions are suppressed.
(ii) Demixing transitions are weakly affected, with small shifts in the $\mu_1-\mu_2$ (chemical 
potentials) plane. (iii) Confinement-induced layering (CIL) transitions occurring in the two 
confined one-component fluids in some cases merge with the demixing transition. 
(iv) Surface-induced layering (SIL) transitions occurring at a single surface as coexistence
conditions are approached are also shifted in the confined fluid. 
Trends with pore size are analysed by means of
complete $\mu_1-\mu_2$ and $p-\bar{x}$ (pressure-mean pore composition) phase diagrams for 
particular values of pore size. This work, which is the first one to address the behaviour
of liquid-crystalline mixtures under confinement, could be relevant as a first step
to understand self-assembling properties of mixtures of metallic nanoparticles under
external fields in restricted geometry.
\end{abstract}

\maketitle

\section{\label{Introduction}Introduction}

Understanding the self-assembly of nanowires and metallic or semiconducting nanorods will be
crucial in next-generation nanoelectronic and display-technology applications \cite{Ahn}. These technologies
exploit the ordering of nanorods to create well-ordered arrays for the selective fabrication of 
devices. Inorganic nanoparticles with controlled size, shape and composition, exhibiting
interesting optical, magnetic and electric properties are now being synthetised \cite{Niidome,Ramanathan}. But most
applications are based on particle arrangements in large structures. Confinement of the rods in 
nanoscale geometries may be one possible strategy to create such structures. 
External electric or magnetic field can be used to orient particles in the required directions \cite{Morrow}.
Predicting the shape and composition effects that lead to ordering in collections of 
nanometer-sized particles in a solvent is a theoretical challenge, and understanding entropic effects 
associated with the rodlike shapes is a prerequisite before considering other interparticle
forces. Therefore, the interest in hard-particle models and in the statistical mechanics of positional
and orientational ordering in fluids of identical or polydisperse hard particles has been revitalised in 
recent years.

The conceptual aspects of freezing of monodisperse particles in nanopores has been studied quite extensively 
\cite{Gubbins,Schoen,Schoen1,Schoen2,Frenkel,Soko,Soko1,Dijkstra,Soko2,Soko3,Dijkstra1,Dijkstra2}. 
On the other hand, smectic and layered phases made of anisometric particles present partial 
(one-dimensional) ordering and give 
an opportunity to study partial, as opposed to complete, spatial order in confined geometries
\cite{Ciach}. Frustration effects in mesophases, normally associated with antagonistic 
conditions imposed on the nematic director in the form of surface and bulk fields, are crucial in technological 
applications of these materials \cite{Frustration}. But in smectic materials another type of frustration
occurs, since the natural layer spacing $d$ establishes a length scale which will compete with any other 
length scale applied in the form of geometric constraints. The simplest confinement geometry is the planar 
slit pore, where the material is sandwiched between two parallel planar substrates. The confinement of 
two- or three-dimensional one-component fluids in the smectic or 
lamellar regime leads to complex behaviour \cite{Polaca,ConfinedSmectic,Yuri}
induced by commensuration effects between the pore
width $h$ and the smectic spacing $d$. Theoretical work indicates that
the possible phase behaviour includes confinement-induced layering (CIL) transitions, as well as a `wavy' 
nematic-smectic transition strongly coupled to the latter. The strong smectic
regime has not been studied experimentally yet.

In contrast to the case of pure materials, confined mixtures of liquid-crystal forming
particles have been given no attention. In our previous work \cite{Nosotros}, we have 
analysed the bulk and adsorption properties of a mixture of hard cylinders.
A complex bulk phase diagram resulted, with nematic (N) and smectic (S) phases.
The smectic phases came in two versions. In the first, non-microsegregated, smectic phase
(S$_1$), layers are identical and composed of a mixture, in various proportions, of the two 
species. In the second, or microsegregated smectic phase (S$_2$), layers rich in one species
alternate with layers rich in the other. Various phase transitions, including first- and 
second-order N-S transitions, as well as smectic demixing, were obtained; see Fig. \ref{fig2}(a),
where the bulk phase diagram is depicted in the plane pressure $p$ vs. composition $x$ (defined as 
mole fraction of the short particles). The adsorption properties of the
mixture on a hard wall were also studied in \cite{Nosotros}. We found a 
complex behaviour, with the wall inducing strong layering leading to stratification of the 
material into alternating layers rich in one of the components. The surface behaviour
includes complete wetting by the S phase, mediated (or not) by an infinite number of surface-induced
layering (SIL) transitions, as well as critical adsorption on approaching second-order N-S transitions. 
As interactions between particles and wall are of hard-core type, layering and wetting properties
of the nematic fluid in contact with the wall have an entropic origin. Similar phenomena have
been found in theoretical work describing the surface phase behaviour of colloid-polymer mixtures
\cite{Colloid}.

In the present paper we extend the analysis of this mixture by considering fluids confined 
by two identical hard walls. This is a novel system where a new length scale appears: apart from the 
pore width $h$, there are two particle lengths, $L_1$ and $L_2$. Commensuration
between the three length scales brings about new phenomena. The most salient results of our work are:
(i) Second-order bulk N-S transitions are suppressed. (ii) Demixing transitions are not much 
affected by confinement, with small shifts in the $\mu_1-\mu_2$ (chemical 
potentials) or $p-\bar{x}$ (pressure vs. mean pore composition) planes. 
(iii) CIL transitions occurring in the one-component fluids are modified
by mixing, ending in critical points or at the demixing transition. (iv) SIL transitions
on each wall are also modified by the confinement; the very first transitions (the lowest-order ones,
involving the first few layers) may survive in wide pores, while the others are preempted
by demixing; in some cases some of them may reappear in the demixing region creating islands of stability,
while higher-order SIL transitions are suppressed. In general, both types of layering transitions 
tend to disappear as the pore width is decreased. (v) Some new phenomena, genuinely connected to 
confinement of the mixture, i.e. to the competition between the three length scales, $L_1$, $L_2$ and
$h$, may also arise.

After summarising the density-functional model used in the analysis and discussing briefly the
bulk and surface phase behaviour in the next section, we present in Section \ref{Confinement}
the results for the confined mixture, considering wide and narrow slit pores. 
Finally, some conclusions are presented in Section \ref{Conclusions}.

\section{\label{ModelBulkSurface}Model, theory and bulk and surface phase diagrams}

\subsection{\label{Model}Model}

\begin{figure}[h]
\includegraphics[width=3.5in]{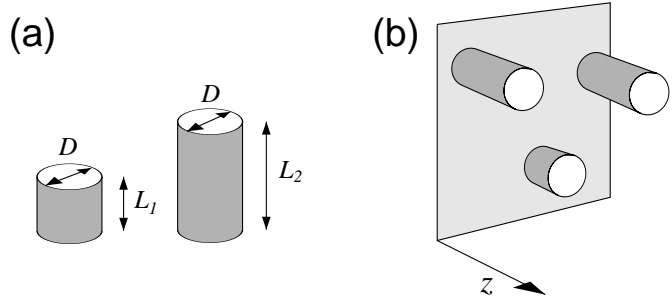}
\caption{\label{fig1} (a) Schematic of particles studied. (b) Geometry of the
adsorbed and confined fluid.}
\end{figure}

We consider a binary mixture of parallel hard cylinders, with lengths
$L_1$, $L_2$ and identical diameters $D$; see Fig. \ref{fig1}(a). 
Label 1 refers to the species of the short
cylinders. We have selected $L_1$ as the unit of length. The length ratio investigated 
is $s=L_2/L_1=3$. {The diameter $D$ is adjusted so that the transverse particle area 
in units of $L_1$ is set to unity, i.e. $\pi D^2/4L_1^2=1$. 
%$L_1=0.89D$, $L_2=2.66D$ (these lengths are adjusted so that the ratio of transverse particle
%area and cylinder length squared of the short species is set to unity, i.e. $\pi D^2/4L_1^2=1$). 
Since particles are parallel, bulk properties only depend on $s$, and the particular values 
of $L_1$ (or $L_2$), and $D$ are irrelevant}. As discussed in \cite{Nosotros}, the values 
of particle lengths
chosen gives a binary mixture that would be equivalent
to a mixture of freely-rotating but highly oriented cylinders (of the same length ratio) such that both components 
would have a stable smectic phase.  The cylinders are in numbers $N_1$ and $N_2$, so that the 
composition of the mixture is defined as $x=N_1/N$, with $N=N_1+N_2$
the total number of particles. The cylinder axes are chosen to lie along the $z$ direction; see
Fig. \ref{fig1}(b). 
This configuration models a real colloidal or molecular fluid where, either by surface
treatment or by means of a bulk field, particles are forced to point along some fixed direction which,
in the case of an adsorption system, would be the surface normal.

\subsection{\label{Theory}Theory}

The properties of the mixture are analysed using a fundamental-measure version of density-functional 
theory \cite{MR1,MR2}. This theory is in very good agreement with Monte Carlo simulations
of monodisperse (one-component) cylinders, and there are no reasons to believe that the theory
is not equally accurate for mixtures. We choose the
grand potential $\Omega$ as the relevant thermodynamic potential. If $\rho_i({\bm r})$
is the local number density distribution of the cylinder centres of mass of the $i$-th species, the 
grand-potential functional of the confined mixture is
\begin{eqnarray}
\Omega[\{\rho_i\}]= {\cal F}[\{\rho_i\}]-\sum_{i=1}^2\mu_i\int_Vd{\bm r}\rho_i({\bm r}),
\end{eqnarray}
where $\mu_i$ is the chemical potential of the $i$-th species. 
The Helmholtz free-energy functional ${\cal F}[\{\rho_i\}]$ is written, as usual \cite{Nosotros}, as
\begin{eqnarray}
{\cal F}[\{\rho_i\}]= {\cal F}_{\rm id}[\{\rho_i\}]+ {\cal F}_{\rm ex}[\{\rho_i\}]+{\cal F}_{\rm ext}[\{\rho_i\}].
\label{eq1}
\end{eqnarray}
${\cal F}$ consists of three parts: the ideal-gas part,
\begin{eqnarray}
{\cal F}_{\rm id}[\{\rho_i\}]=
kT\sum_{i=1}^2\int_V d{\bm r} \rho_i({\bm r})\left\{\log{[\cal V}_i\rho_i({\bm r})]-1\right\},
\end{eqnarray}
the excess part,
\begin{eqnarray}
{\cal F}_{\rm ex}[\{\rho_i\}]=kT\int_V d{\bm r} \Phi(\{\rho_i({\bm r})\}),
\end{eqnarray}
and the external contribution from the walls:
\begin{eqnarray}
{\cal F}_{\rm ext}[\{\rho_i\}]
=kT\sum_{i=1}^2\int_V d{\bm r}\rho_i({\bm r})V_{\rm ext}^{(i)}({\bm r}).
\end{eqnarray}
Note that temperature $T$ is not a relevant variable since all interactions are purely hard and
therefore ${\cal F}$ is proportional to $T$. In the above $k$ is Boltzmann constant, ${\cal V}_i$ is the
thermal volume of species $i$, and $\Phi(\{\rho_i({\bm r})\})$ is the excess free-energy density. 
{In the fundamental-measure approximation \cite{MR1,MR2}, 
$\Phi({\bf r})$ depends on the local averaged one-body $n_i({\bm r})$ and two-body 
weighted densities $N_i({\bf r})$. 
However, by imposing the density profiles to be functions only on $z$ (smectic symmetry), these 
weighted densities reduce to only two (one-body) densities, $n(z)$ and $\eta(z)$. In this case
the expression for the free-energy density is given by}
\begin{eqnarray}
\Phi=n\left\{-\ln\left(1-\eta\right)
+\frac{3\eta}{1-\eta}+\frac{\eta^2}{\left(1-\eta\right)^2}
\right\}.
\label{fmt}
\end{eqnarray}
The averaged densities are
\begin{eqnarray}
n(z)=\frac{1}{2}\sum_{i=1}^2
\left[\rho_i\left(z-\frac{L_i}{2}\right)+\rho_i\left(z+\frac{L_i}{2}\right)\right],
\hspace{0.6cm}\eta(z)=a_0\sum_{i=1}^2\int_{z-L_i/2}^{z+L_i/2}\rho_i(z')dz',
\label{n3}
\end{eqnarray}
where $a_0=\pi D^2/4$ is the transverse area of the cylinder. $\eta(z)$ is the local
packing fraction. 

{The procedure to obtain the expression for $\Phi({\bf r})$ consists of the following 
two steps: (i) a density functional is derived for a mixture of two-dimensional hard disks, with 
the important property of dimensional cross-over from two to one dimension (i.e. when 
disks are located on a straight line, the functional should reduce to the exact one for a mixture of hard 
segments). (ii) If we impose the parallel-alignment constraint on particles, the density functional for 
a mixture of parallel hard cylinders can be obtained from the density functional
for the hard-disk mixture just derived by applying again the dimensional cross-over criterion. This 
is easy to visualize by taking into account that the projections of the
cylinders on a plane perpendicular to their axes just give
a mixture of disks with different diameters \cite{MR1}.}

%In the above expressions, use has been made of the symmetry of our 
%problem: both in bulk (smectic phase) or in inhomogeneous situations
%(adsorption or confined geometries), density distributions will only depend on the $z$ coordinate. 
{In the interfacial problems to be studied below, inhomogeneities resulting from the presence of
surfaces will be included via an external potential, which consists of hard walls:}
\begin{eqnarray}
V_{\rm ext}^{(i)}({\bm r})=
\left\{\begin{array}{ll}\displaystyle\infty,&z\displaystyle<\frac{L_i}{2}\hspace{0.2cm}
\hbox{or}\hspace{0.2cm}z>h-\frac{L_i}{2},\\\\
\displaystyle 0,&\displaystyle\frac{L_i}{2}\le z\le h-\frac{L_i}{2},\end{array}\right.\hspace{0.6cm}
i=1,2,
\label{Vext}
\end{eqnarray}
where $h$ is the width of the slit pore. The conditions on the confined fluid are controlled by the chemical
potentials $\mu_1$ and $\mu_2$ of the bulk phase with which it is in equilibrium. Equivalently
the bulk phase can be characterised by the pressure $p$ and the bulk composition $x$ (in case of
demixing in the bulk phase the transformation $(\mu_1,\mu_2)\to(p,x)$ is not unique with respect
to the composition variable). The bulk fluid mixture and the adsorption problem on a single
hard wall (semiinfinite system) can be studied within the same scheme \cite{Nosotros}.
From the density profiles one can define the partial mean densities $\bar{\rho}_i$, the
total mean density $\bar{\rho}$ and the global mole fraction $\bar{x}$, as:
\begin{eqnarray}
\bar{\rho}_i=\frac{1}{h}\int_0^hdz \rho_i(z),\hspace{0.4cm}\bar{\rho}=\bar{\rho}_1+\bar{\rho}_2,\hspace{0.4cm}
\bar{x}=\frac{\bar{\rho}_1}{\bar{\rho}}.
\end{eqnarray}
In Sec. \ref{Confinement} we will present phase diagrams of the confined mixture in
the $\mu_1$-$\mu_2$ plane but also in the $p$-$\bar{x}$ plane, since in the latter 
demixing regions can be appreciated. For fixed $\mu_1$, $\mu_2$, as $h\to\infty$ we have $\bar{x}\to x$.
The equilibrium density profiles can also be obtained in other ensembles, and we have
found it convenient in some cases to use the Gibbs free energy per particle $G_N=G/N$ which depends,
in particular,
on the mean pore composition $\bar{x}$. $G_N$ allows the study of fractionation effects (gaps in
$\bar{x}$) at first-order transitions. This can be carried out by minimization of $G_N$ with respect to
$\bar{\rho}$ and $x_i(z)\equiv \rho_i(z)/\bar{\rho}$ for a fixed value of mean pore
composition $\bar{x}$. The Gibbs free energy per particle can be obtained
from the Helmholtz free energy per unit volume ${\cal F}_V\equiv{\cal F}/V$ by a Legendre
transformation, $G_N=\left({\cal F}_V+\Gamma\right)/\bar{\rho}$, where
$\displaystyle{\Gamma=-\frac{\partial\left({\cal F}_V/\bar{\rho}\right)}{\partial \bar{\rho}^{-1}}
=-\Omega[\{\rho_i^{(\rm{eq)}}\}]/V}$, i.e. minus the grand potential per unit volume evaluated at
the equilibrium density profiles \cite{cuenta}. Thus we obtain the usual result $\Gamma=p$, the bulk pressure, when
external potentials are absent. The double-tangent construction on the function $G_N(\bar{x})$ for
a fixed value of $\Gamma$ is equivalent to the equality of chemical potentials of different
species in each of the coexisting phases. These chemical potentials in turn coincide with those
appearing in the definition of $\Omega[\rho]$. Changing the value of $\Gamma$, and repeating the above procedure,
we can calculate the phase diagram in the coordinates $p$--$\bar{x}$. Note that the bulk pressure
$p$ can be calculated from the values of $\{\mu_i\}$.

In the following we summarise the bulk behaviour and the surface behaviour of the fluid
against a single wall.

\subsection{\label{Bulk}Bulk phase behaviour}

\begin{figure}
\includegraphics[width=5.4in]{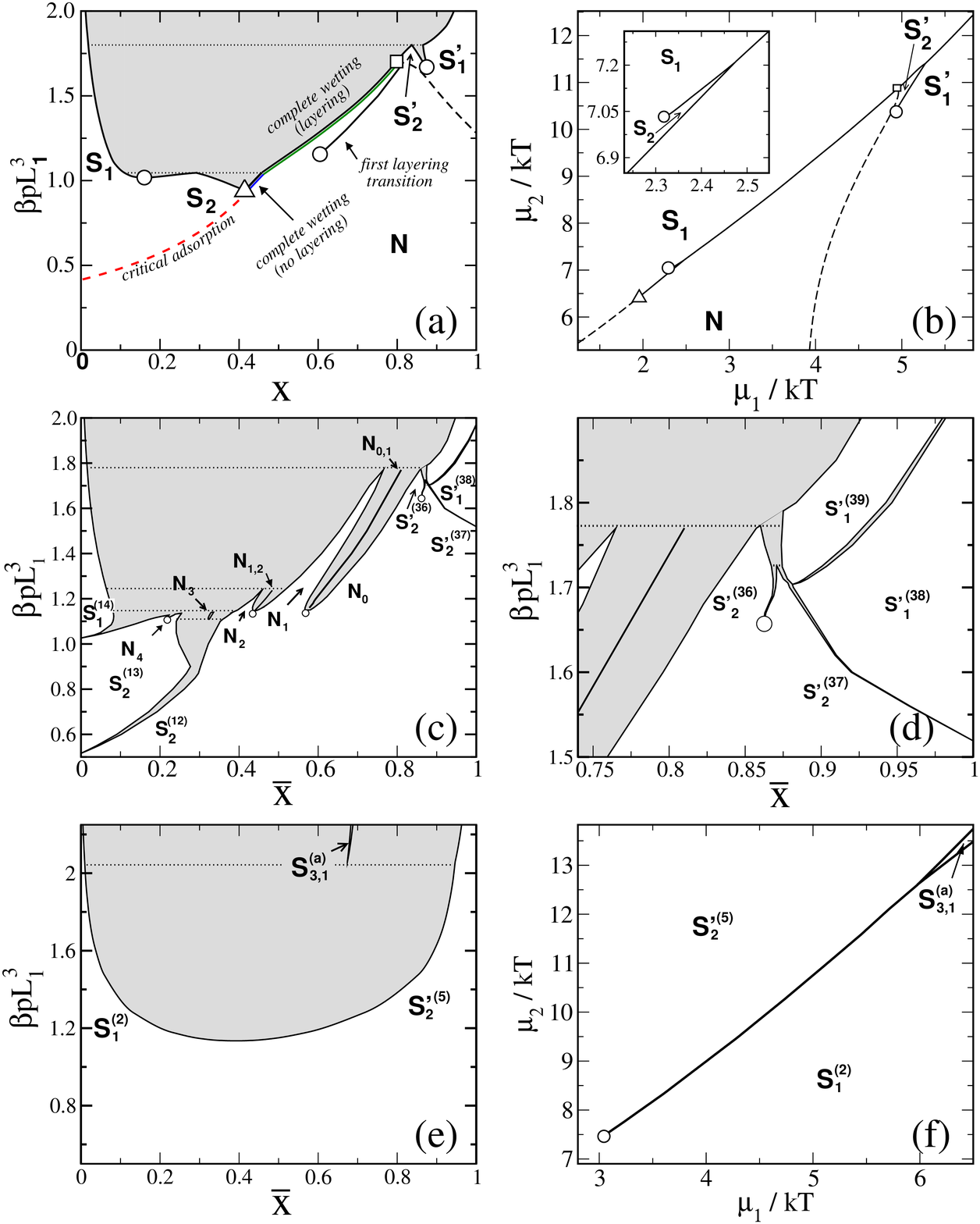}
\caption{\label{fig2} 
(a) Bulk phase diagram of the mixture in the $p$-$x$ plane. Text indicates type of wetting behaviour.
(b) Bulk phase diagram of the mixture in the $\mu_1$-$\mu_2$ plane. 
(c) Phase diagram of the confined mixture in the $p$-$\bar{x}$ plane for pore width $h=48.2 L_1$. 
(d) Zoom of the high $p$ and $\bar{x}$ region in panel (c).
(e) Phase diagram of the confined mixture in the $p$-$\bar{x}$ plane for pore width $h=6.8 L_1$. 
(f) Phase diagram of the confined mixture in the $\mu_1$-$\mu_2$ plane for pore width $h=6.8 L_1$.
In all diagrams circles indicate critical points, triangles tricritical points, squares
critical end points; dashed curves are continuous phase transitions. Dotted horizontal lines indicate 
triple or quadruple points.}
\end{figure}

In \cite{Nosotros} we analysed the bulk phases of the above model mixture. Here we briefly
discuss the main results. For the nematic phase
$\rho_i(z)=\rho_i=$const., whereas in the smectic phase $\rho_i(z)$ are periodic functions
of $z$, with period $d$. The stability of the nematic and smectic phases was obtained in a 
$p$-$x$ (pressure-composition) phase diagram \cite{Nosotros} (details of these calculations 
can be found therein). Fig. \ref{fig2}(a) presents the bulk phase diagram in
this plane; in part (b) the same diagram is shown in the $\mu_1$-$\mu_2$ plane.
The following features are worth mentioning: (i) there are two second-order
N-S transitions (dashed lines), each starting at one of the pure-fluid cases, $x=0$ or $x=1$; 
(ii) one of these transitions, the one coming from the $x=0$ axis, is connected to a 
first-order N-S transition by a tricritical point [triangle in Fig. \ref{fig2}(a)]; and (iii) 
three regions of S-S demixing, two of them ending in corresponding critical points (circles), appear in
the phase diagram. Note that, in the regions of smectic stability,
two smectic structures occur: one where the density waves of the two species are in
phase (one-layer smectic), denoted by S$_1$, and another where the densities are out of phase 
(two-layer smectic), S$_2$. The latter presents a microsegregation of the two species. By changing the
conditions on the mixture one can pass from one of these smectic phases to the other in a continuous
fashion. In the bulk phase diagrams of Figs. 
\ref{fig2}(a) and (b), and in the rest of the article, the smectic phases are denoted by primed
or unprimed labels according to whether the smectic is rich or poor in the short component,
respectively.

{We now briefly comment on the expected changes in phase diagram topology if particles 
axes were allowed to rotate. The most important difference will be the presence of the isotropic phase,
and the existence of an associated first-order isotropic-nematic phase transition with a transition gap that
depends on the mixture composition. Obviously our model, with the perfect-alignment constraint,
cannot account for this transition. Also, our model predicts a second-order nematic-smectic transition
which, as show below, is suppressed by confinement (an effect ultimately arising from the 
perfect-alignment approximation). However, in situations where the bulk phases (nematic or smectic)
are highly oriented, we expect a phase behavior very similar to that shown in the present work.}  

\subsection{\label{Wetting}Surface behaviour}

The behaviour of the single-wall system is very important to understand the 
properties of the confined mixture, as will be seen in the following section.
The surface behaviour of the fluid when a nematic phase is adsorbed 
on a substrate was investigated in our previous work \cite{Nosotros}. Particle axes were taken to point along
the normal to the substrate, which was represented by an external potential
$V_{\rm ext}({\bm r})$ of the same type as in (\ref{Vext}). We only 
studied the case where the bulk fluid in equilibrium with the surface structure was nematic.
The nematic phase far from the wall $(z\to\infty)$ was characterised by values of the pressure
$p$ and the composition $x$ or, equivalently, by the chemical potentials of the two species,
$\mu_1$ and $\mu_2$. A conjugate-gradient scheme was used in the numerical minimisation (see details in Ref.
\cite{Nosotros}), from which equilibrium density distributions and values of surface
tensions were obtained.

\begin{figure}
\includegraphics[width=3.6in]{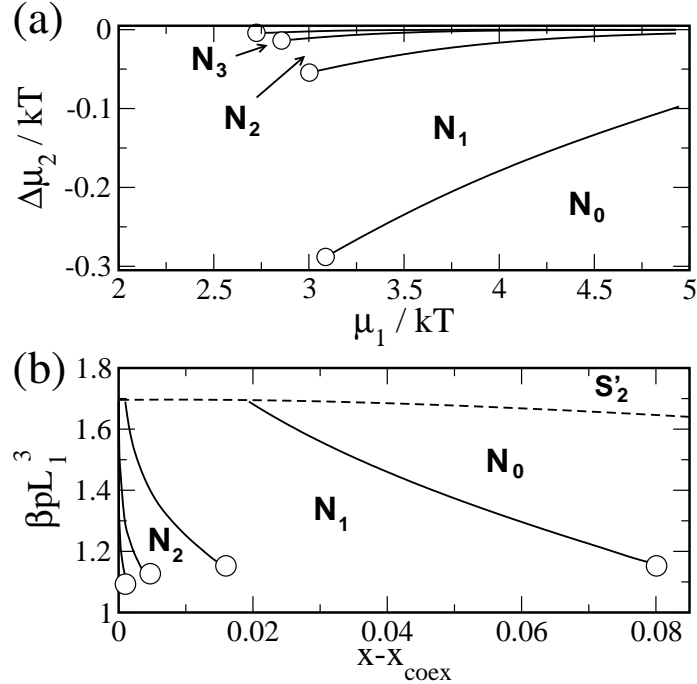}
\caption{\label{nueva2} SIL transitions in the planes (a) $\mu_2$-$\mu_1$, and (b) $p$-$x$ in
the single-wall system. In (a), $\Delta\mu_2$ is the chemical potential of species 2 with
respect to the value of the bulk nematic-smectic transition. In (b), $x_{\rm coex}$ is the nematic 
value at the transition for each value of pressure.}
\end{figure}

At a given value of pressure $p$, wetting of the wall by the smectic phase was found 
for all pressures as $x\to x_{\rm NS}(p)$, where $x_{\rm NS}(p)$ is the boundary of the
N-S transition. However, depending on the bulk pressure, three r\'egimes may occur: 
(i) Complete wetting by the S$_1$ phase, rich in long particles, via an infinite sequence of 
{\it surface-induced layering} (SIL) transitions when $p$ is above the S$_1$-S$_2$-N triple point; at a 
first-order SIL transition, the composition of a localised interfacial region, with a thickness of 
one layer, changes abruptly from a low to a high value, as a result of which the short component is 
expelled from the wall and sharply localised, smectic-like layers, adsorb at the wall in a 
stepwise fashion, until a macroscopically thick smectic S$_1$ film develops at the
wall (complete wetting). (ii) Complete wetting by the S$_2$ phase without SIL 
transitions when $p$ is below the triple point but above the tricritical point. And (iii) critical 
adsorption at the second-order bulk N-S$_2$ transition, i.e. for pressures below the tricritical point.
The first of the infinite sequence of SIL transitions in the first r\'egime has been plotted
in Fig. \ref{fig2}(a); the rest are too close to the coexistence boundary to be seen. A blow up
of the first four transitions is shown in Fig. \ref{nueva2} in two different planes. Here, and in the
rest of the paper, phases denoted by N$_n$ refer to structures with $n$ localised layers of the long
particles near the wall, and a density that tends to a constant value far from the wall; in the confined case
the density may not be completely uniform in the middle of the pore, but the structures are
generally connected with the bulk nematic as $h\to\infty$.

\section{\label{Confinement}Confinement}

One of the most striking phenomena presented by the mixture of cylinders is the occurrence of SIL
transitions at a single wall as the bulk N-S coexistence is approached. Our explanation for
this phenomenon \cite{Nosotros} is that there is an entropic competition (reduction of
excluded volume) between the two species to cover the wall, which couples to translational
entropy to produce a first-order phase transition. But there is another type of layering phenomena,
i.e. the {\it confinement-induced layering} (CIL) transitions, which take place when a spatially ordered 
phase competes with a confining length scale. These phenomena are well documented and have been 
investigated theoretically in liquid crystals and related 
materials \cite{Polaca,ConfinedSmectic}. CIL phenomena should also happen in fluid mixtures that 
can stabilise a smectic phase in bulk. Since the smectic phase of our mixture 
comes in two varieties (S$_1$ and S$_2$) and the bulk phase diagram is already quite complex,
we can anticipate a fascinatingly rich phase diagram when the mixture is confined and the
possible occurrence of new phases.

\begin{figure}
\includegraphics[width=3.6in]{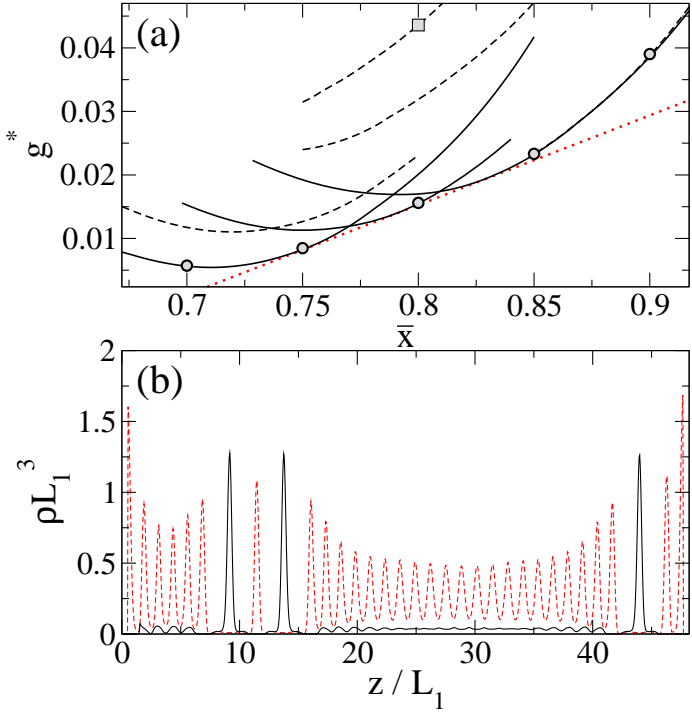}
\caption{\label{gx} 
(a) Gibbs free-energy density $g^*=G_N/kT$ as a function of global mole
fraction $\bar{x}$ for the case $\beta pL_1^3=1.7$. For the sake of visualisation, a straight
line $-10.96+6.00\bar{x}$ has been subtracted from the free energies.
Curves correspond to minima at fixed $\bar{x}$ obtained from
the conjugate-gradient method (depending on the initial conditions different branches can be obtained).
Continuous curves: branches that eventually lead to stable states. Dashed curves: metastable branches.
Circles correspond to states obtained by simulated annealing. Dotted (straight) 
line indicates 
three-phase coexistence of a confined SIL transition. (b) Profiles of the metastable state
indicated by the square in panel (a); the continuous curve is the profile of the long particles, whereas 
the dashed curve corresponds to the short particles.}
\end{figure}

As explained above, the particle model used in our calculations exhibits a second-order bulk N-S
transition. Due to suppression of fluctuations along the $z$ direction when the fluid is confined,
the N-S transition disappears in the confined system. Therefore, it is not possible to assign in a clear
manner the roles of nematic and smectic to any of the confined phases, except in some particular
cases (generally speaking, as pressure or chemical potential is increased, the oscillating distribution of particles 
tends to become sharper throughout the pore). The particular cases are: (i) Confined phases close to
$x\agt 0$ and $x\alt 1$ existing in a closed region of stability, which are generally connected with the bulk smectic
phase as $h\to\infty$. (ii) Surface phases in the confined case, which undergo SIL transitions, and which are
connected with N$_n$-type phases in the single-wall ($h=\infty$) case (in bulk, these phases are in contact with
a bulk nematic phase, but in the confined case the latter may have more or less spatial order depending on
the values of pore width and pressure). Our convention is to denote all confined phases by S$^{(m)}$, with $m$
the number of layers of either species inside the pore, in case (i), and by N$_n$, with $n$ the number of absorbed 
layers of the long particles, in case (ii). Intermediate cases are not well defined by these notations.

\begin{figure}
\includegraphics[width=3.6in]{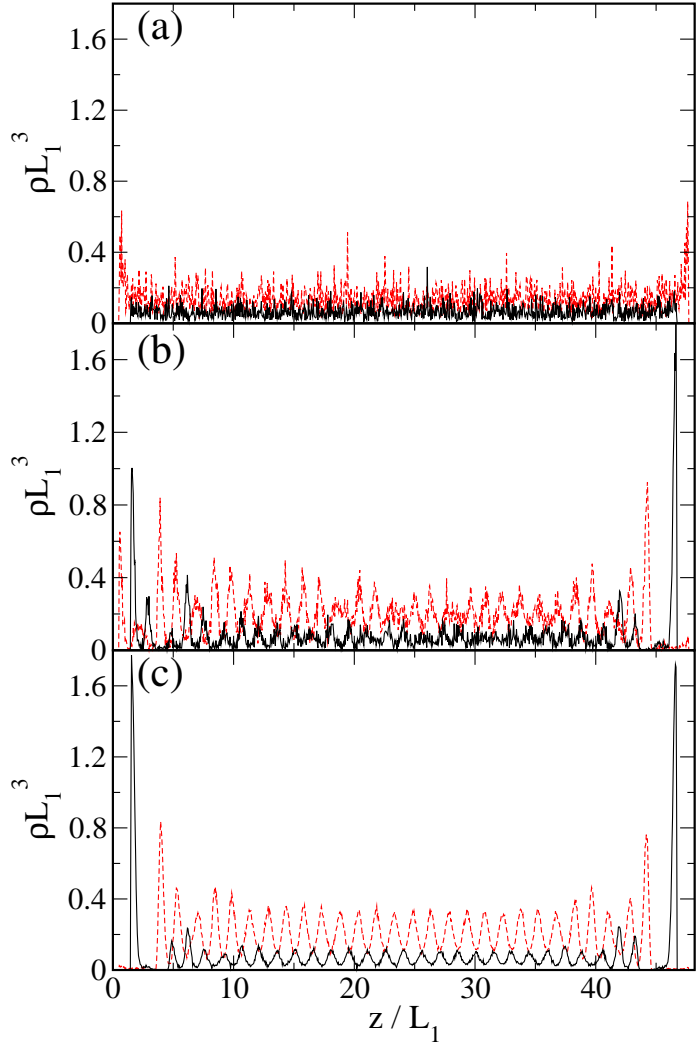}
\caption{\label{SA} From (a) to (c), sequence of profiles in a simulated-annealing simulation
for the case $\beta pL_1^3=1.7$, $\bar{x}=0.7$ and $h=48.2L_1$. The continuous curves are
the profiles of the long particles, whereas the dashed curves correspond to the short particles. 
A complete animation of the simulation can be found in \cite{Animation}.}
\end{figure}

In the confined case the method of solution of our theory 
is as follows. For fixed values of $\mu_1$, $\mu_2$, the fluid equilibrium
structure within the pore was obtained using a conjugate-gradient (CG) method, where the $z$ coordinate is
discretised as $z_k=k\Delta z$, with $\Delta z=0.01L_1$, and $\rho_k^{(i)}\equiv\rho_i(z_k)$ are used as
discrete variables. However, the free-energy landscape of the confined mixture is very complex, with
many local minima of similar free energy, corresponding to different arrangements of
particles that can accommodate into the pore. The grand-potential minimum given by the CG method 
depends crucially on the initial configuration used to start the iterative CG algorithm. Alternatively,
we have used a simulated-annealing (SA) method to double-check that the minima found are actually the
absolute minima. Even then, one always has to bear in mind that one cannot be completely certain as
to the absolute nature of the minima found, especially in particular areas of the phase diagram 
where the particle arrangements may be more degenerate. This is the case 
in the region of the phase diagram where the values of $p$ and 
$\bar{x}$ are high, i.e. in a compressed mixture where most of the particles are short.
Fig. \ref{gx}(a) shows the structure of minima in the Gibbs free-energy landscape as a function of 
$\bar{x}$ for the case $\beta pL_1^3=1.7$.
Curves are minima obtained using the CG method, starting from different initial 
profiles. Depending on the initial condition, the minimisation procedure gets trapped into one of
many possible metastable branches. Panel (b) of the figure shows a typical metastable state [actually
the one located by the square in panel (a)]. For a given value of $\bar{x}$, the different metastable branches 
correspond to the very many different ways of distributing the layers of the long species. Circles correspond to 
SA simulations starting from uniform profiles which hopefully, and in the cases explored it was always so, 
are able to identify the correct branch. 
In the case shown, there exists coexistence between three structures with different values of $\bar{x}$, which follows 
from the double-tangent construction, for a SIL transition. 
Fig. \ref{SA} shows the evolution of a SA simulation for the case $\beta pL_1^3=1.7$, $\bar{x}=0.7$ and
$h=48.2L_1$.
The SA procedure is able to find the minimum free-energy configuration in a few hundred steps \cite{Animation}.
The resulting structure is then refined using the CG algorithm.
{Although the difference in free energy between the coexisting phases may be relatively small
(as is the case shown in Fig. \ref{gx}), the values of the composition variable $\bar{x}$ 
are not so similar (see values at coexistence in the figure). This in turn means that the coexisting phases
have different interfacial structures. Therefore, in case this scenario were to
occur in an experimental system, large energy barriers separating these states may
be expected, so that only
fluctuations with relatively large amplitudes can take the system out from its original
state. This scenario may be typical when large and moderate demixing gaps separate two
stable confined phases. However, as the pore compositions of two coexisting phases become
similar, and consequently also their interfacial structures become similar, these fluctuations,
present in the real experimental system but not taken into account in the present model,
can suppress these transitions. 
Inclusion of fluctuations in the orientations of the particle axes might stabilize
interfacial structures such as those shown in Fig. \ref{gx} (b), which is due to the lowering 
of the elastic energy when particles rotate in such a way as to modify the period of the confined smectic
and better commensurate the number of smectic layers with the pore width.}

In order to understand the results for the confined mixture, let us
first discuss the effects of confinement on the one-component 
fluid. The bulk fluid possesses a second-order transition from
the nematic to the smectic, which disappears in the confined system due to
the finite size available for spatial correlations. The only 
feature of the phase diagram that remains on confinement is the 
presence of first-order CIL transitions. These 
transitions occur at high values of pressure or chemical potential, i.e.
when the smectic order inside the pore is well developed. As mentioned 
already, their origin is the fact that, for a given value of pore width $h$, 
the confined fluid can only accommodate a particular number $n$ of smectic 
layers, which is given approximately by $h\simeq nd$, where 
$d$ is the equilibrium layer spacing of the bulk smectic.
In general the confined smectic will be stressed, either because it is 
compressed or swollen. As $h$ is varied, the number of layers will be
increased or decreased to optimise the available space from a thermodynamic
point of view. Layering transitions are mostly associated with variations
in $h$, and depend more weakly on pressure or chemical potential, i.e.
they are almost vertical in a $p$-$h$ or $\mu$-$h$ phase diagram. CIL transitions 
terminate at critical points (see Ref. \cite{ConfinedSmectic}).
There is a terminal, lower pore width at which layering transitions
cease to exist.

\begin{figure}
\includegraphics[width=3.6in]{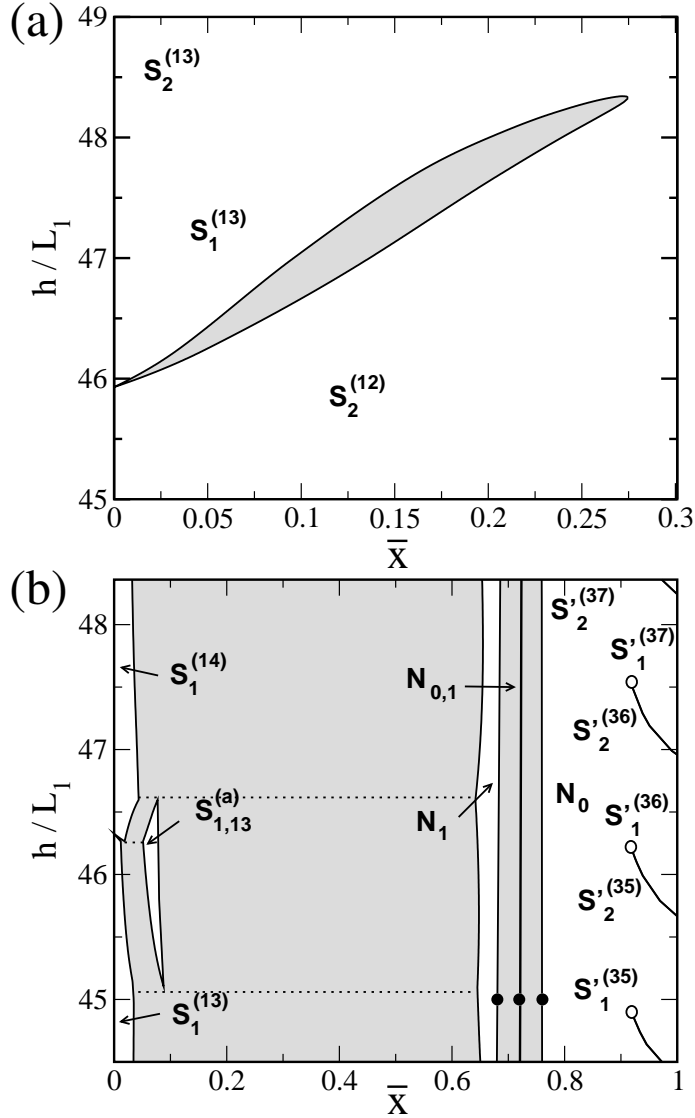}
\caption{\label{fig3} Phase diagrams of the confined mixture in the $h$-$x$ plane for two values of the
pressure. (a) $\beta pL_1^3=0.81$. (b) $\beta pL_1^3=1.51$. Labels indicate nature of phases.
Horizontal dotted lines denote triple points.}
\end{figure}

\subsection{Wide slit pores}

Starting from the one-component fluid, as one adds a small amount of particles 
of the other component, the structure of CIL transitions is shifted in the
phase diagram. If short particles are added to a smectic structure made of long
particles, the smectic spacing will tend to increase and we expect the CIL
transitions to occur at larger values of $h$. The distance between two consecutive 
transitions will be more or less constant. An example of this phenomenon is
presented in Fig. \ref{fig3}(a), which shows a CIL transition between smectic
phases S$_2^{(12)}$ and S$_1^{(13)}$, containing 12 and 13 layers, respectively,
in a mixture at low pressure ($\beta pL_1^3=0.81$). 
Note that there is a mole-fraction gap at the transition, i.e. the global mole 
fractions $\bar{x}$ of the two coexisting structures are different. There is also 
a terminal pore width beyond which the layering transition disappears. Above 
the transition gap in Fig. \ref{fig3}(a) the coexisting phase S$_1^{(13)}$ smoothly
changes to an out-of-phase structure S$_2^{(13)}$. On the other side of the
phase diagram ($x\alt 1$), CIL transitions are not occurring at this value of the
pressure since the mixture is in the N bulk regime, but they would certainly occur at
higher pressure.

\begin{figure}
\includegraphics[width=3.6in]{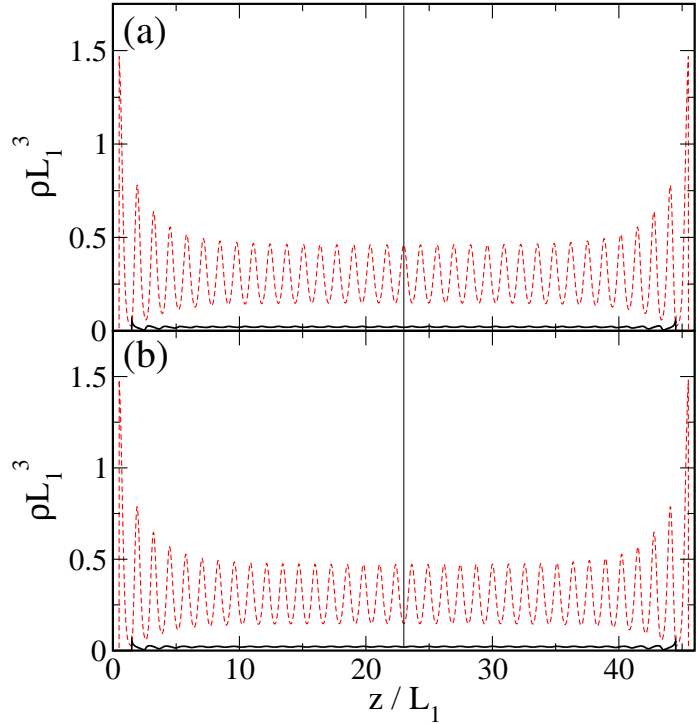}
\caption{\label{nueva1} Density profiles of two phases that coexist at a CIL transition in the
confined fluid with $\beta pL_1^3=1.51$ and $h=46.0 L_1$ [see Fig. \ref{fig3}(b)]. (a)
S$_2^{\prime(35)}$ phase, with $\bar{x}=0.9385$. (b) S$_1^{\prime(36)}$ phase, with
$\bar{x}=0.9391$. The thin vertical line divides the slit pore in two identical parts.
Continuous curve: species 2 (long particles). Dashed curve: species 1 (short particles).}
\end{figure}

At higher pressures the phenomenology becomes much more complex, as CIL transitions
compete with demixing. This was to be expected, as the bulk phase diagram
shows that demixing occurs for pressures $\beta pL_1^3\agt 1.0$. An example is
shown in Fig. \ref{fig3}(b). For the pressure shown, $\beta pL_1^3=1.51$,
strong N-S demixing occurs at bulk [Fig. \ref{fig2}(a)], and correspondingly
the confined mixture also exhibits demixing in a similar mole-fraction interval.
But with important differences, which will be commented on later. First note that,
in the range $x\alt 1$, smectic CIL transitions of the same type as in the region
$x\agt 0$ for the lower pressure do exist. They are separated approximately by a
pore-width increment $\Delta h\simeq 1.2L_1\simeq d$, the bulk smectic
spacing of the short species. Fig. \ref{nueva1} shows the density profiles of two
phases that coexist at the S$_2^{\prime{\rm (35)}}$-S$_1^{\prime{\rm (36)}}$ CIL transition.

\begin{figure}
\includegraphics[width=3.6in]{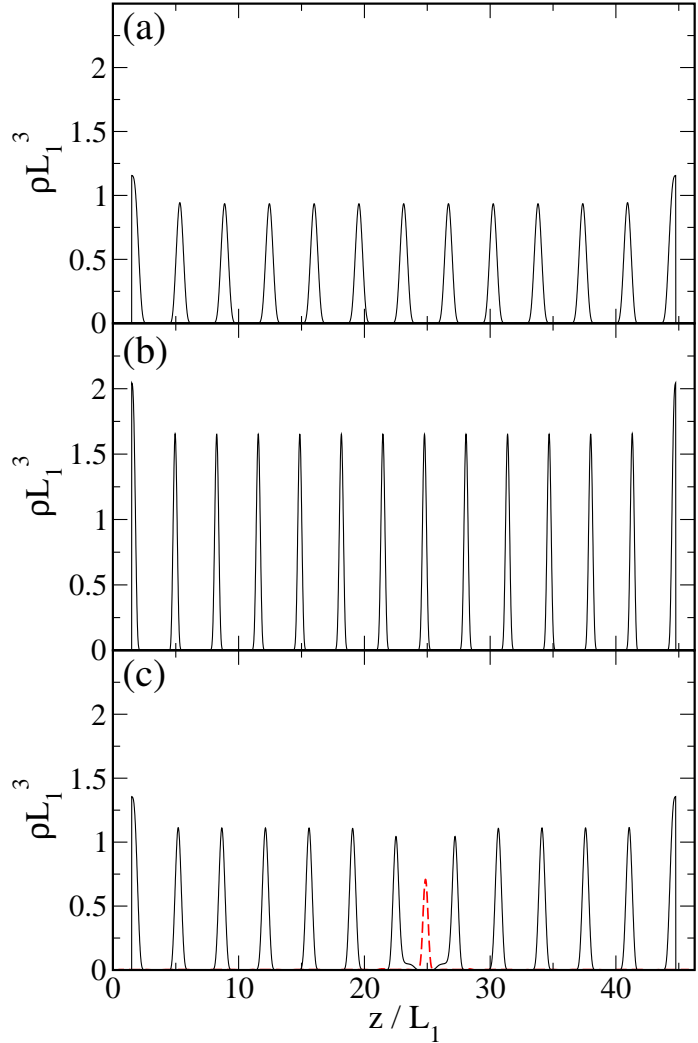}
\caption{\label{fig4} Density profiles of the three phases that coexist at a value of pressure
$\beta pL_1^3=1.51$ and for a pore width $h=46.3L_1$ [see Fig. \ref{fig3}(b)].
(a) S$_1^{(13)}$ phase, with $\bar{x}=0.012$. (b) S$_1^{(14)}$ phase, with $\bar{x}=0.019$. (c) S$_{1,13}^{(\rm a)}$, with
$\bar{x}=0.053$. Continuous curve: species 2 (long particles). Dashed curve: species 1 (short particles).
Note that, in panels (a) and (b), the density of short particles
is too small to be visible in the graphs.}
\end{figure}

In the region $\bar{x}\agt 0$, CIL transitions are rapidly preempted by a demixing 
instability. One interesting feature of the phase diagram of Fig. \ref{fig3}(b) is that, in the proximity of 
the CIL transition, an asymmetric structure, denoted by S$_{1,13}^{(\rm a)}$, is
stabilised in a small island in the $h$-$\bar{x}$ plane; presumably, these islands exist associated with
other CIL transitions, creating phases S$_{n,m}^{(\rm a)}$ (in this notation, $n$ and $m$ are the number
of layers of species 1 and 2 of the asymmetric structure, respectively). In the case shown, the island interacts 
with the CIL and demixing transitions, creating triple points where three confined fluids coexist. 
For the S$_1^{(13)}$-S$_1^{(14)}$ CIL transition, the asymmetric structure,
S$_{1,13}^{(\rm a)}$, contains 7 smectic layers of long particles at one wall, 
separated from 6 smectic layers of long particles at the other wall, by a single localised
layer of the (minority) short species. Note that other very similar structures with almost identical
values of free energy but with different positions of the single peak of the short particles may exist,
but structures with more than one layer of short particles,
for larger values of $\bar{x}$, are unstable with respect to demixing.
The triple points are indicated by horizontal dotted lines in Fig. \ref{fig3}(b), while
the density profiles of the three phases that coexist at the intermediate triple point, involving the
S$_1^{(13)}$, S$_1^{(14)}$ and S$_{1,13}^{(\rm a)}$ phases,
can be seen in Fig. \ref{fig4}. 

\begin{figure}
\includegraphics[width=3.6in]{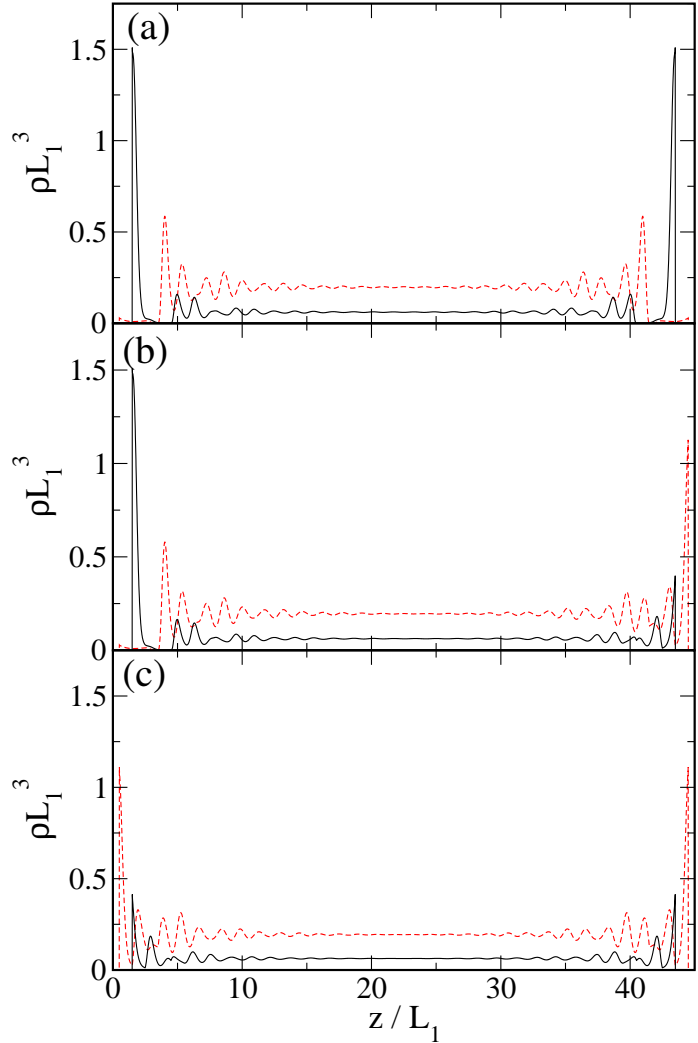}
\caption{\label{fig5} Density profiles of the three phases that coexist at a value of pressure
$\beta pL_1^3=1.51$ and for a pore width $h=45.0L_1$ [see Fig. \ref{fig3}(b)].
(a) N$_1$phase, with $\bar{x}=0.688$. (b) N$_{0,1}$ phase, with $\bar{x}=0.724$. (c) N$_0$ phase, with
$\bar{x}=0.759$. Continuous curve: species 2 (long particles). Dashed curve: species 1 (short particles).}
\end{figure}

Leaving aside these islands of stability within the demixing region, we can see
that this region consists of two segregation `bands': a first, wide band, in the
mole-fraction interval $0.04\alt\bar{x}\alt 0.65$, and a second, narrow band, in the
interval $0.68\alt\bar{x}\alt 0.76$. The first band corresponds to the shifted bulk
demixing N-S transition, while the second is associated with the SIL transitions
occurring in the single-wall system, but here in the confined case. 
Inspection of Fig. \ref{fig5} indicates that
this is indeed the case. In the figure, density profiles that coexist at a pressure
$\beta pL_1^3=1.51$ in a pore of width $h=46.3L_1$ are shown [filled circles in 
Fig. \ref{fig3}(b)]. Profiles in panels (a) and (c) are symmetric and correspond
to the phases with low and high values of mole fraction, respectively. The structural differences
between the two profiles are identical to those occurring at a single-wall interfacial
structure undergoing a SIL transition, described in Sec. \ref{Wetting}.
The mole-fraction gap at this SIL transition arising in the confined fluid tends to
decrease as $h$ is decreased, until it eventually disappears at a critical point,
not shown in Fig. \ref{fig3}(b). The transition is connected to a corresponding SIL transition at a single
surface as $h\to\infty$. Inside the instability region of the confined 
SIL transition an almost vertical, continuous line has been drawn, which corresponds to a peculiar
structure that coexists with the other two [panel (b) in Fig. \ref{fig5}].
This structure, which has interfacial structures identical to that of (a) at the
left wall and to that of (b) at the right wall, is asymmetric but has the same
interfacial free energy as the other two (an equivalent structure is obtained
by interchanging the right and left interfacial profiles). The existence of structure
(c) is the consequence of this phenomenon being associated to the SIL transitions,
which in the confined case, and for wide pores, occur at both surfaces independently
without any interaction. As the pore width is reduced, interaction between the two 
surfaces structures would make the line inside the instability region in Fig. \ref{fig3}(b) to have a finite 
width. This structure seems too elusive to be detected experimentally, since 
domains of the two coexisting interfacial structures would probably mix in different proportions
at both surfaces.

\begin{figure}
\includegraphics[width=3.6in]{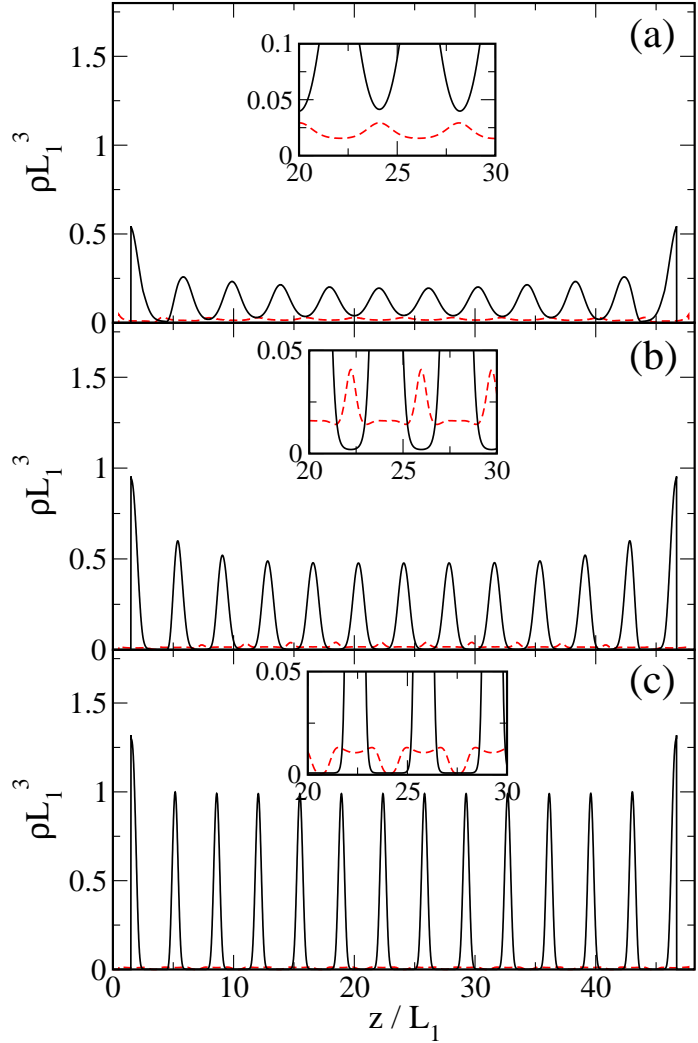}
\caption{\label{fig6} 
Density profiles of three representative states labelled in Fig. \ref{fig2}(c) as 
S$_2^{(12)}$, S$_2^{(13)}$ and S$_1^{(14)}$ for a pore width $h=48.2L_1$.
(a) S$_2^{(12)}$ phase: $\beta pL_1^3=0.6$, $\bar{x}=0.14$; (b)
S$_2^{(13)}$ phase: $\beta pL_1^3=1.0$, $\bar{x}=0.10$; and (c) S$_1^{(14)}$ phase: 
$\beta pL_1^3=1.2$, $\bar{x}=0.05$. Continuous curves: species 2. Dashed lines: species 1.}
\end{figure}

The $p$-$\bar{x}$ phase diagram of the confined mixture for a particular value of pore width,
$h=48.2 L_1$ (in the regime of wide pores) is plotted in Fig. \ref{fig2}(c). The diagram contains
the same features as previously described, but viewed in a different way. Also, it allows to
visualise critical points associated with confined SIL transitions and some additional features.
The diagram is to be compared with the corresponding bulk diagram. The wide demixing region is 
quite similar. But some differences are apparent. First, as mentioned before,
the second-order bulk N-S transition is vanished. Second, the structure of CIL
transitions appears in the regions $x\agt 0$ and $x\alt 1$, and is seen to
either interact with demixing or terminate in critical points. 
Since $h$ is fixed, the number of transitions in the phase diagram 
is very limited (CIL transitions occur mainly with respect to pore width and are weakly 
dependent on pressure). In Fig. \ref{fig6} the density profiles of three representative 
structures associated with the CIL transitions are shown. These phases are labelled as
S$_2^{(12)}$, S$_2^{(13)}$ and S$_1^{(14)}$. In the first two, the density profiles
of the two species are out of phase, whereas in the third the minority (short) species
adopts an in-phase configuration with two small, separated peaks about the main smectic
layer; in this case the short particles go into the main layer, forming a wide two-in-one
layer where short particles flow more or less freely.

An important feature in the phase diagram of Fig. \ref{fig2}(c) is that the remnant of some of the 
(infinitely many) SIL transitions occurring in the single-wall case are clearly apparent. For the pore width
shown, $h=48.2 L_1$, only four of them survive. The first one, involving the phases N$_0$-N$_1$,
exhibits a long segregation region ending in a critical point; this structure is reminiscent
of the corresponding N$_0$-N$_1$ transition in the single-wall system, and in the variable $\bar{x}$
the transition shows a gap. The second transition, N$_1$-N$_2$, is much shorter but exhibits the same
characteristics. The third transition occurs right in the middle of the demixing region and, instead
of being a separated segregation region, it appears as a very small stability island [N$_3$ in Fig. \ref{fig2}(c)]. 
Finally, the N$_3$-N$_4$ transition is very short, and again presents an associated critical
point. As mentioned in the discussion on the $h$-$\bar{x}$ phase diagram, Fig. \ref{fig3}(b), 
asymmetric phases appear in the central region of the mole-fraction gap in the case of the first
two SIL transitions; these structures are indicated by the labels N$_{0,1}$ and N$_{1,2}$ in Fig. \ref{fig2}(c).

\begin{figure}
\includegraphics[width=3.6in]{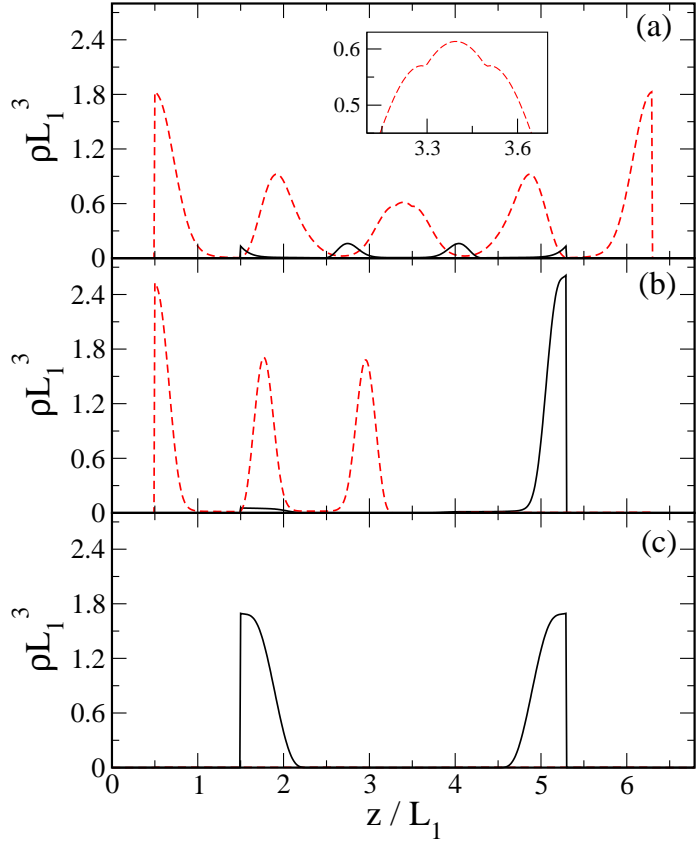}
\caption{\label{fig7} 
Density profiles of three states that coexist at $\beta pL_1^3=2.0$ with pore
width $h=6.8 L_1$ [see Fig. \ref{fig2}(e)]. (a) S$_2^{\prime(5)}$ phase, with $\bar{x}=0.944$; 
(b) S$_{3,1}^{(a)}$ phase, with $\bar{x}=0.674$, and
(c) S$_1^{(2)}$ phase, with $\bar{x}=0.010$. Continuous curves: species 2. Dashed curves: species 1.}
\end{figure}

Finally, we focus on the region at the upper-right corner of the phase diagram of Fig. \ref{fig2}(c).
This region is quite complex. A zoom of this region is presented in Fig. \ref{fig2}(d). The small demixing region
separating the bulk S$_1^{\prime}$ and S$_2^{\prime}$ phases, with the corresponding critical point,
still survives for this pore width. Again, the bulk second-order N-S transition disappears, but
the CIL transitions associated with the one-component short-particle fluid survive in mixtures
with up to 12\% of long particles. These transitions interact with the small demixing region.
A final feature in Fig. \ref{fig2}(d) is the existence of a quadruple point, indicated by the horizontal
dotted line.

\subsection{Narrow slit pores}

In the case of narrow pores, available space is much more limited and much of the rich phenomenology
described for wide pore vanishes. We have investigated the case $h=6.8 L_1$. The $p$-$\bar{x}$
phase diagram is depicted in Fig. \ref{fig2}(e). A large demixing region persists, but the diagram
is otherwise relatively featureless: both SIL and CIL transitions seem to have disappeared, and
the only feature that remains is
a very narrow stability region, deep in the demixing region, where a largely asymmetric structure
is stabilised. As a result, a new triple point arises. This asymmetric structure might be the remnant
of a SIL transition in wider pores. Fig. \ref{fig7} shows the density profiles of the three phases
that coexist at the triple point. The phase shown in panel (b) may be interpreted as a demixed
phase where mixing entropy is compensated by an optimised surface entropy. 
Fig. \ref{fig2}(f) is the corresponding phase diagram in the $\mu_1$-$\mu_2$ plane. The demixing
transition is slightly shifted with respect to the corresponding bulk transition [see Fig. \ref{fig2}(b)].

\begin{figure}
\includegraphics[width=5.6in]{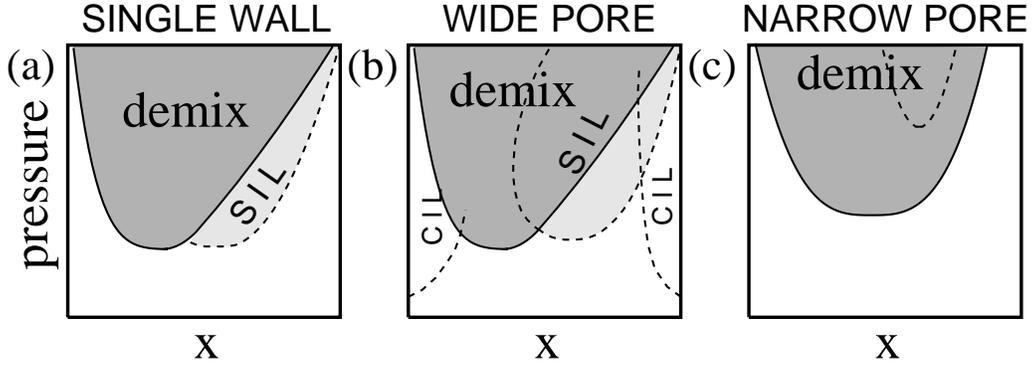}
\caption{\label{sum} Schematic of phase diagram evolution as the slit pore width is reduced.}
\end{figure}

\subsection{Kelvin equation}

{The macroscopic approach to capillary transitions is given by the Kelvin equation,
which relates the degree of under- (or over-) saturation of the capillary phase transition, with 
respect to a reservoir, in say chemical potential, $\Delta\mu$, to the surface tension 
$\gamma_{\alpha\beta}$ of the interface between the two coexisting phases $\alpha$ and $\beta$, 
the contact angle $\Theta$ and the pore width $h$
(in the case of a wetting layer, the pore width has to be diminished by twice the thickness of
the wetting layer $w$). The approach is only valid asymptotically, in the regime where $h\to\infty$.
For a mixture, the Kelvin equation has been discussed by Evans and
Marini Bettolo Marconi \cite{Umberto}. We can apply it to our particular capillary demixing system.
Let $\alpha$ denote the phase favoured by the walls, in our case the $S_1$ phase, and $\beta$
the phase present in the central region of the pore, in our case N. We should remind ourselves
that the real coexisting pore structure, N$_n$, consists of two `smectic' films adsorbed on the
walls with $n$ layers each and a central,
more or less uniform, nematic film (the value of $n$ depends on $p$ and $h$. In a very wide pore,
$n$ will be large and it makes sense to split the N$_n$ structure into two adsorbed films and
a central one; in a narrower pore, the smectic layers in N$_n$ may be just a few in number, and such a
division, and as a consequence Kelvin equation, is not valid). Then the shift in chemical
potential $\Delta\mu_2$ of the long species (more abundant in the $\alpha$ phase) with respect
to the bulk value is given by \cite{Umberto}
\begin{eqnarray} 
\Delta\mu_2=\frac{2\gamma_{\alpha\beta}B^{-1}\cos{\Theta}}{h\rho_1^{(\alpha)}},
\end{eqnarray} 
where the $B$ coefficient is
\begin{eqnarray} 
B=\left(\frac{1-x}{x}\right)_{\alpha}-\left(\frac{1-x}{x}\right)_{\beta}
=\frac{x_{\beta}-x_{\alpha}}{x_{\alpha}x_{\beta}}.
\end{eqnarray} 
and where the composition values are given at bulk. Since we have wetting at bulk coexistence,
$\cos{\Theta}=1$ and $h\to h-2w$. To check the validity of the equation in our present setup,
we have applied the equation for a pressure $\beta pL_1^3=1.25$
and for the wider pore width used in this work, $hL_1^{-1}=48.2$ (in the case of the narrower pore,
$hL_1^{-1}=6.8$, the whole procedure is meaningless, as there is no way to identify smectic and 
nematic films in the fluid structure, since the pore is too narrow). We have $w\approx 2L_2=6$
[for $\beta pL_1^3=1.25$ we have coexistence between the S$_1$ and N$_2$ phases, see Fig. \ref{fig2}(c)],
$x_{\alpha}=0.0588$, $x_{\beta}=0.5823$, $\beta\gamma_{\alpha\beta}L_1^2=0.055451$, and
$\rho_1^{(\alpha)}L_1^3=0.0103$, which gives $B^{-1}=0.0654$ and $\beta\Delta\mu_2=0.019$. This is
to be compared with $\beta\Delta\mu_2=0.029$ from the density-functional calculation. The discrepancy
is a huge 35\%, and the reason is twofold: (i) the pore width $h=48.2L_1$ only spans
approximately 15 lengths of the long particle; this is too small for the Kelvin equation to be
accurate. (ii) In confined layered phases, such as the smectic phase, there occur important 
commensuration effects between the intrinsic periodicity of the phase and the pore width, and
an elastic contribution, containing the layer compressibility modulus, to the Kelvin equation 
must be included \cite{Ciach,ConfinedSmectic}. 
For wide pores the simple Kelvin equation would give predictions 
for the capillary phase transition that are roughly averages over the real values (which are 
oscillatory with respect to the pore width, with a period of approximately one particle length).}

\section{\label{Conclusions}Conclusions}

\begin{figure}
\includegraphics[width=4.0in]{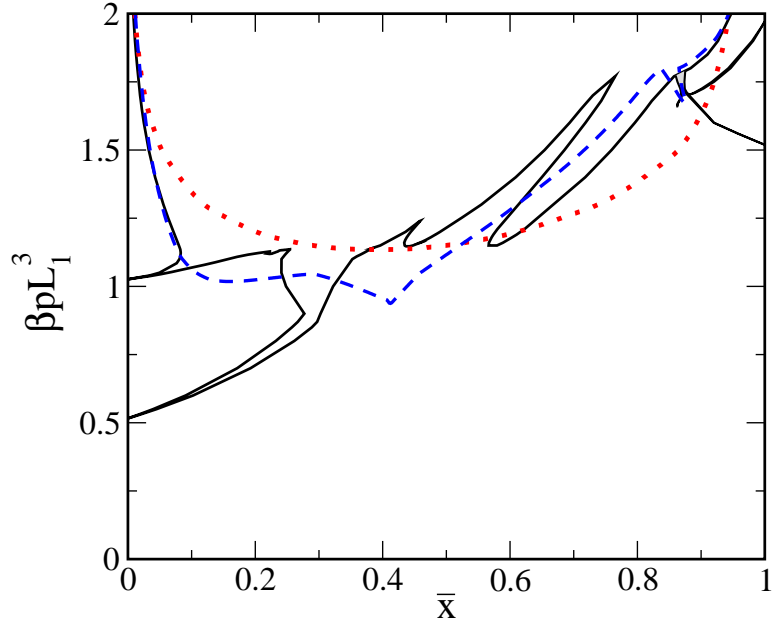}
\caption{\label{comparison} Demixing regions in the plane $p$-$\bar{x}$ for the three cases
investigated in this work: $h=\infty$ (dashed curve), $h=6.8L_1$ (dotted curved) and
$h=48.2L_1$ (continuous curve).}
\end{figure}

In this paper we have discussed the phenomenology of a confined mixture of hard cylindrical
particles oriented in the direction perpendicular to the confining hard walls. Competition
between the pore width $h$ and the two lengths of the particles, $L_1$ and $L_2$, create
a complex behaviour. Three regimes can be identified, see Fig. \ref{sum}. The first is the single-wall 
case, $h=\infty$ [Fig. \ref{sum}(a)], where there is a wide demixing region and SIL transitions induced by 
the wall-fluid interaction; these off-coexistence transitions are associated with a regime of
complete wetting by a smectic phase when nematic conditions prevail at bulk. In the 
second regime, that of wide but finite pores, $L_1,L_2\ll h<\infty$ [Fig. \ref{sum}(b)], a new feature 
arises: CIL transitions, induced by confinement, when the composition of the mixture is close to
zero or unity (i.e. almost one-component cases), which may or may not interact with a 
wide demixing region. SIL transitions are affected by the confinement, and only a finite 
number of them survive. Finally, in the regime of narrow pores, i.e. wall separations only slightly
larger than both particle lengths, $h>L_1,L_2$ [Fig. \ref{sum}(c)], only
demixing survives, and weak signatures of the layering transitions may be seen in the
form of narrow stability islands in the demixing mole-fraction gap. As expected for mixtures of
particles with such different volumes, demixing is always quite strong. The segregation
region shows a surprisingly invariant mole-fraction gap, and quite similar values for the
pressure above which demixing occurs, with respect to pore width. This is reflected in Fig. 
\ref{comparison}, where the demixing regions for the three cases investigated in this paper have been 
superimposed.

This work has focused on confined phases with smectic symmetry. The possible stability of the columnar
phase is a question that is left for future work. A complete calculation of the absolute stability of the 
columnar phase in the context of the present model is a difficult task which will have to
be tackled using numerical techniques different from the ones developed here. In our previous
work \cite{Nosotros} we used a bifurcation analysis to show that the nematic-columnar spinodal
line is always below the nematic-smectic line for all values of composition. This is not a proof that 
the smectic phases are stable with respect to the columnar phase, but at least indicates that part of the 
phases obtained here could be stable in some range of pressures.

\acknowledgements

We acknowledge support from the Direcci\'on 
General de Universidades e Investigaci\'on of the Comunidad de
Madrid (Spain), under the R\&D Programmes of activities MODELICO-CM/S2009ESP-1691 and
NANOFLUID, and to the Ministerio de Educaci\'on y Ciencia of Spain (grants
FIS2007-65869-C03-01, FIS2008-05865-C02-02 and MOSAICO).


\begin{references}
\bibitem{cuenta} This result can be easily obtained from the chain rule
$\Gamma=\bar{\rho}\partial {\cal F}_V/\partial\bar{\rho}-{\cal F}_V=
V^{-1}\left[\bar{\rho}\sum_i\int dz\left(\delta{\cal F}/\delta\rho_i(z)\right)
\left(\partial\rho_i(z)/\partial \bar{\rho}\right)-{\cal F}\right]$, from the definition
$\bar{\rho}\partial\rho_i(z)
/\partial\bar{\rho}=\bar{\rho}x_i(z)=\rho_i(z)$ and from the equilibrium
conditions $\delta {\cal F}/\delta \rho_i(z)=
\mu_i$ to be applied when $\rho_i(z)\neq 0$, resulting in $\Gamma=V^{-1}\left[\sum_i\mu_i\int dz\rho_i(z)-{\cal F}\right]
=-\Omega/V$.
\end{references}
\end{document}